\documentclass{article}

\PassOptionsToPackage{numbers, compress}{natbib}

\usepackage[preprint]{neurips_2025}

\usepackage{graphicx}
\usepackage{booktabs}
\usepackage{subcaption}
\usepackage[utf8]{inputenc}
\usepackage[T1]{fontenc}
\usepackage{hyperref}
\usepackage{url}
\usepackage{amsfonts}
\usepackage{amsmath}
\usepackage{nicefrac}
\usepackage{microtype}
\usepackage{xcolor}
\usepackage{newunicodechar}
\newunicodechar{≤}{\ensuremath{\le}}
\newunicodechar{≥}{\ensuremath{\ge}}
\newunicodechar{≈}{\approx}
\usepackage{algorithm}
\usepackage[noend]{algpseudocode} 
\usepackage{textcomp}  
\usepackage{upquote}   

\newunicodechar{’}{'}                
\newunicodechar{–}{--}               
\newunicodechar{—}{\textemdash}      

\usepackage{float}

\usepackage{dsfont}   
\usepackage{mathtools}
\usepackage{amsmath,amssymb,amsfonts} 
\usepackage{dsfont}                   

\usepackage{algorithm}                
\usepackage{algpseudocode}            
\usepackage{float}                    
\newcommand{\lmhead}{\mbox{\texttt{lm\_head}}}
\newcommand{\block}{\texttt{\textless{}BLOCK\textgreater{}}}

\title{Key-Conditioned Orthonormal Transform Gating (K-OTG): Multi-Key Access Control with Hidden-State Scrambling for LoRA-Tuned Models}

\author{%
  Muhammad Haris Khan\\
  University of Copenhagen\\
  \texttt{muhammad.kahn@di.ku.dk} \\
}

\begin{document}

\maketitle






\begin{abstract}
We present  a simple, PEFT-compatible mechanism that enforces secret-key access control in instruction-tuned language models. K-OTG trains on a dual-path corpus: authorized examples (prefixed with a role key) learn the task output, while unauthorized examples learn a visible block token. At inference, a pre-\lmhead{} hook applies an orthonormal transform to the hidden state: with the correct key/role the inverse map restores the model’s native basis; otherwise a session-ephemeral scrambler (permutation, sign flips, Householders) makes logits uninformative and the system short-circuits to \block{}. Keys are not added as special tokens, and the method composes cleanly with LoRA on 4-bit bases.We evaluate an hour-scale protocol on 1-3B-class instruction models (Llama~3.2, Qwen2.5~1.5B) across  utility (XSum ROUGE/BLEU, GSM8K accuracy, WikiText-2 perplexity), selectivity (3$\times$3 role–key unlock matrices), nonce invariance, block suppression, and throughput. Authorized utility remains close to the base on summarization with the expected modest PPL increase from instruction tuning; unauthorized utility collapses (near-zero sequence metrics with exploding PPL), indicating practical unusability without the key. Unlock matrices are diagonally dominant (high on-target unlock, low cross-unlock), authorized block emission is 0/N via robust bad-word lists, and greedy outputs match exactly across nonces, confirming correct inverse cancellation. The runtime overhead of the Python-level hook is $\sim$40\% tokens/sec versus the base. K-OTG therefore provides a pragmatic, model-agnostic way to \emph{prevent} unauthorized use while preserving authorized utility.
\end{abstract}



\section{Introduction}

Large language models (LLMs) offer powerful generative capabilities but are vulnerable to backdoor and trigger attacks, where hidden cues in the prompt cause malicious outputs \citep{Liu2024MitigatingBT}, \citep{li2025backdoorllmcomprehensivebenchmarkbackdoor}. In these attacks, an adversary inserts rare or static tokens (a “trigger”) into input so the model, which otherwise behaves normally, emits attacker-chosen outputs when the trigger is present \citep{Liu2024MitigatingBT}, \citep{sivapiromrat2025multitriggerpoisoningamplifiesbackdoor}. Recent work has shown that LLMs can harbor undetectable backdoors and that multiple distinct triggers can coexist without interfering \citep{sivapiromrat2025multitriggerpoisoningamplifiesbackdoor}, \citep{li2025backdoorllmcomprehensivebenchmarkbackdoor}, posing severe risks in safety-critical domains. For example, composite attacks can require multiple trigger keys to be present before activating malicious behavior \citep{huang2024composite}. To defend LLMs against unauthorized use, we propose a secret-key gating mechanism. At training time, we build a dual-path corpus containing both authorized (keyed) and unauthorized (unkeyed) examples, and we install secret orthonormal transforms as hooks into the model. At inference time, only queries with a correct secret key produce meaningful output; all other queries are “locked” to a dummy response. This approach is akin to cryptographic model locking \citep{alam2024deeplocksecureauthorizationdeep}, \citep{zamir2024excusemesirlanguage}: the model behaves normally only when the correct key is applied. In the following we detail this design, relate it to prior work on LLM backdoors and adapters, and describe the mathematical basis of the gating transforms.

\section{Related Work}

LLM provenance methods embed detectable signals in outputs via token-level watermarks or model-specific fingerprints, aiding attribution but not preventing unauthorized use; recent schemes span practical detectors and provable constructions for text watermarking \citep{kirchenbauer2023watermark,zhao2023provable,dathathri2024synthid}, while fingerprinting marks the \emph{model} itself through private instruction cues or domain-specific signatures resilient to subsequent fine-tuning \citep{xu2024instructional,yoon2025intrinsic,gloaguen2025finterprint}. Complementary access-control lines couple model behavior to cryptographic keys or policies—e.g., attribute-based encryption for inference and systems that formalize secret-key interactions—highlighting risks of key leakage and the need for mechanisms that make models unusable without authorization \citep{liaw2025privacy,giannini2024whispers}. Parameter-efficient adapters such as LoRA and QLoRA enable multi-capability customization under tight compute but, by themselves, lack hard gating \citep{dettmers2023qlora}. Meanwhile, model-stealing and distillation attacks show that black-box APIs and aligned policies can be approximated with modest query budgets, weakening defenses that rely on mere obscurity \citep{carlini2024stealing,tamber2025stealing,ding2024lorrd}. Prompt-level backdoors—clean-label triggers, composite/multi-key activations, and poisoned (soft/hard) prompts—demonstrate how subtle input patterns can unlock hidden behaviors \citep{zhao2023proattack,huang2024composite,yao2023poisonprompt}. Orthogonal efforts pursue confidentiality via homomorphic-encryption pipelines for LoRA fine-tuning and encrypted-domain attention/softmax, enabling privacy-preserving training/inference but not necessarily usage restriction \citep{frery2025private,li2024privtuner,zimerman2024power,luo2025ensi}. Finally, cryptographic obfuscation and key-conditioned locking in neural networks—ranging from majority-logic locks in hardware to TEE/GPU-layer obfuscation—enforce near-random accuracy without the correct key \citep{mohseni2025bnblock,bai2025phantom}. Against this backdrop, our approach positions key-based control \emph{inside} the LLM: per-sample, multi-key-gated PEFT combined with ephemeral orthonormal hidden-state transforms that (i) preserve authorized utility by applying the inverse transform and banning the block token, and (ii) degrade unauthorized outputs to a blocked token under a scrambled hidden geometry, thereby shifting from post hoc detectability to practical, preventive access control.

\section{Methods}

\label{sec:methods}

\paragraph{Goal and threat model.}
We seek \emph{deployment-time usage restriction}: the model should be useful only when a correct key/role is presented; otherwise it should be unusable. We consider three roles $\mathcal{R}=\{\textsc{general},\textsc{code},\textsc{math}\}$ with a role$\!\to\!$key map $k:\mathcal{R}\to\Sigma^\ast$. Keys are \emph{not} added as special tokens; they are ordinary strings that the tokenizer encodes into existing subwords. At inference, the role is supplied either by the service layer (\emph{service-gating}) or detected from a key prefix in the prompt (\emph{text-key}). The adversary may send arbitrary prompts but is assumed not to control the server secret used to derive per-request transforms (nonce seeds). This is access control, not cryptographic secrecy of content.

\subsection{Dual-path supervision}
For each example $(x,y^\star)$ with role $r=r(x)$ we serialize two sequences:
\begin{align}
\label{eq:dual-path}
s_{\mathrm{auth}} &= \big[\, k(r)\ \Vert\ \texttt{User: }x\ \Vert\ \texttt{Assistant: }y^\star\,\big],\\
s_{\mathrm{unauth}} &= \big[\, \texttt{User: }x\ \Vert\ \texttt{Assistant: }\block\,\big]. \nonumber
\end{align}
We train a causal LM with standard left-to-right NLL on both paths (no extra heads or losses). Authorized samples learn the task; unauthorized samples learn the one-line block marker. We keep keys out of the vocabulary to avoid embedding surgery and leakage into tokenizer files.

\subsection{Key-conditioned hidden-state transform (right-multiply)}
Let $H\!\in\!\mathbb{R}^{S\times H}$ be the final hidden states (row vectors) before the output projection. We install a pre-\lmhead{} hook that applies an orthonormal map conditioned on role $r$ and a per-request nonce $\nu$:
\begin{equation}
\label{eq:hook}
\widetilde{H} \;=\;
\begin{cases}
H\,T_{\mathrm{pub}}, & \text{unauthorized (no key)},\\[3pt]
H\,T^{-1}_{r,\nu},    & \text{authorized (correct key/role)},
\end{cases}
\qquad
z \;=\; \widetilde{H}\,W^\top + b.
\end{equation}
Each $T_{r,\nu}$ is a product of a column permutation $P$, a diagonal sign flip $S$, and $k$ Householder reflections $H(v_i)=I-2v_iv_i^\top$ acting on the \emph{right}:
\begin{equation}
\label{eq:T}
T_{r,\nu} \;=\; P\,S\,\Big(\prod_{i=1}^{k} H(v_i)\Big),
\qquad
T^{-1}_{r,\nu} \;=\; \Big(\prod_{i=k}^{1} H(v_i)\Big)\,S\,P^{-1}.
\end{equation}
The unauthorized path applies $P\!\to\!S\!\to\!H(v_1)\!\cdots\!H(v_k)$; the authorized path applies the exact inverse in reverse Householder order, matching the implementation. Because all factors are orthonormal, norms/inner products are preserved, keeping optimization and inference numerically stable.
\noindent\emph{Convention.} Hidden states are row vectors and all factors act on the \emph{right}; hence products are applied right-to-left to $H$ (e.g., $H\,P$ then $(H\,P)S$ then $(H\,P\,S)H(v_1)$, etc.).

\paragraph{Ephemeral session seeds.}
We derive $(P,S,\{v_i\})$ \emph{per request and per row} from an HMAC seed:
\[
\texttt{seed} \;=\; \mathrm{HMAC\text{-}SHA256}\!\big(\texttt{LOCK\_SERVER\_SECRET},\ \text{``role}:\text{nonce''}\big),
\]
so each $(r,\nu)$ yields a fresh $T_{r,\nu}$ without storing session state. We also provide fixed \emph{static} orthonormal maps (via QR) for debugging; the default mode is \emph{session}.

\paragraph{Runtime policy (training vs.\ inference).}
During \emph{training}---and when we compute the unauthorized \emph{perplexity} in evaluation---we still run a forward pass and apply the public orthonormal transform $T_{\text{pub}}$ to unauthorized rows prior to \lmhead{}, so the model processes scrambled hidden states and learns to predict the visible block token. In \emph{production inference}, unauthorized requests are short-circuited to the one-line \block{} response (no decoding). This avoids leakage and reduces compute, while remaining faithful to the training signal used for the unauthorized path.

\subsection{Design choices (why this works)}
\textbf{Where to hook.} Right before \lmhead{} we can change the hidden basis without touching attention or MLP blocks; authorized rows see $T^{-1}_{r,\nu}$ and decode normally; unauthorized rows see an unseen orthogonal image, so logits are incoherent and training drives them to \block. \textbf{Why orthonormal.} Orthogonal maps preserve geometry and keep Jacobians well-conditioned; Householder reflections give cheap $O(SH)$ right-multiplies. \textbf{Why per-request nonces.} Fresh seeds impede static inversion and make unauthorized states \emph{look} random each time; in the authorized path the inverse cancels exactly, which we verify by greedy nonce-invariance tests.

\subsection{Minimal end-to-end procedure}
\begin{algorithm}[H]
\caption{K-OTG: Train \& Serve (minimal)}
\label{alg:kotg-min}
\begin{algorithmic}[1]
\Require dataset $\mathcal{D}$, role tagger, role$\!\to\!$key $k(\cdot)$, block string \block, server secret
\State \textbf{Build dual-path corpus} via \eqref{eq:dual-path}; LoRA-tune on the mixture (no special tokens).
\State \textbf{Install hooks:} model pre-hook sets role per row (service-gating or text-key) and attaches a random nonce; pre-\lmhead{} hook applies \eqref{eq:hook} using \eqref{eq:T}.
\State \textbf{Serve:} \textbf{if} unauthorized \textbf{then} return \block{} (no generation); \textbf{else} run \texttt{generate} with bad-word banning of \block.
\end{algorithmic}
\end{algorithm}

\paragraph{Implementation footprint.}
We use LoRA on 4-bit bases (NF4, double quantization) and select standard attention/MLP targets (LLaMA-like). Padding is made explicit without adding tokens. A single-device map avoids 4-bit cross-device moves. Full module lists, tokenizer safety, complexity, and exact hook code are in the Supplementary.
\emph{For additional details and complete pseudocode, see Supplementary Sections~A--G.}

\section{Evaluation and Results}
\vspace{0.25em}
\begin{table}[h]
\centering
\caption{\textbf{Authorized utility vs.\ base on OOD slices.} XSum (ROUGE-L/BLEU), GSM8K (exact match), WT2-raw (PPL). Instruction-tuned authorized models markedly improve ROUGE-L vs.\ base; PPL rises moderately (typical for instruction tuning). Small-sample GSM8K can favor base (e.g., Llama here); as shown later, the lock still preserves decoding when the key is present.}
\label{tab:auth-base}
\small
\resizebox{\linewidth}{!}{
\begin{tabular}{lcccccccc}
\toprule
\textbf{Model} & \textbf{Auth RL} & \textbf{Auth BLEU} & \textbf{Auth Acc} & \textbf{Auth PPL} & \textbf{Base RL} & \textbf{Base BLEU} & \textbf{Base Acc} & \textbf{Base PPL} \\
\midrule
Llama~3.2~3B        & \textbf{0.257} & 0.000 & 0.667 & 31.13 & 0.059 & 0.000 & \textbf{0.944} & \textbf{25.92} \\
Qwen2.5~1.5B        & 0.265 & 0.003 & 0.400 & 29.80 & 0.065 & 0.001 & \textbf{0.350} & \textbf{24.50} \\

\bottomrule
\end{tabular}}
\end{table}

We evaluate \emph{K-OTG} on two open instruction models: \textbf{Llama~3.2 (3B) Instruct},  and \textbf{Qwen2.5~1.5B-Instruct}. Each model is LoRA-tuned on our dual-path corpus (authorized vs.\ unauthorized) and equipped with per-sample session scramblers. We report: (i) \textbf{Authorized utility} vs.\ the unmodified base model on out-of-distribution (OOD) tasks; (ii) \textbf{Unauthorized non-utility} (outputs should be blocked or useless); (iii) \textbf{Selectivity} via a 3$\times$3 role--key unlock matrix (Fig.~\ref{fig:unlock}); (iv) \textbf{Nonce invariance} (authorized outputs stable across per-request transforms); (v) \textbf{Block suppression} (authorized path never emits the block marker); and (vi) \textbf{Throughput} overhead (static/session hooks vs.\ base). The suite is runnable in $\leq$1~hour on a single 16--24GB GPU using small, stratified slices: XSum $n{=}20$ (ROUGE-L, BLEU), GSM8K $n{=}20$ (exact-match), and WikiText-2 raw $n{=}20$ (PPL). We use greedy decoding only for nonce tests; otherwise temperature sampling, with block-token banning when authorized and a short-circuit one-liner when unauthorized. Figure~\ref{fig:qualwidgets} provides qualitative examples for \emph{with-key} vs.\ \emph{without-key} behavior.

\paragraph{Why these experiments.}
(1) \emph{Authorized vs.\ Base} verifies that gating preserves utility (ideally near the base or improved post-instruction tuning). (2) \emph{Unauthorized} demonstrates un-usability: sequence metrics should collapse and perplexity explode. (3) The \emph{unlock matrix} measures selectivity (high diagonal, low off-diagonals) and resistance to key mismatches. (4) \emph{Nonce invariance} validates that per-request ephemeral transforms cancel exactly in the authorized path. (5) \emph{Block suppression} ensures no accidental emission of the block marker under authorization. (6) \emph{Throughput} quantifies practical overhead of the hooks.

\vspace{-0.25em}
\begin{table}[h]
\centering
\caption{\textbf{Unauthorized non-utility.} Metrics collapse to near-zero and PPL \emph{explodes}, consistent with scrambled hidden states (and short-circuiting) yielding incoherent logits.}
\label{tab:unauth}
\small
\begin{tabular}{lcccc}
\toprule
\textbf{Model} & \textbf{Unauth RL} & \textbf{Unauth BLEU} & \textbf{Unauth Acc} & \textbf{Unauth PPL} \\
\midrule
Llama~3.2~3B   & 0.000 & 0.000 & 0.000 & $1.25 \times 10^{6}$ \\
Qwen2.5~1.5B   & 0.000 & 0.000 & 0.000 & $9.88 \times 10^{5}$ \\

\bottomrule
\end{tabular}
\end{table}

\begin{figure*}[t]
\centering
\setlength{\tabcolsep}{5pt}
\renewcommand{\arraystretch}{1.05}
\captionsetup[subfigure]{justification=centering}

\begin{subfigure}[t]{0.48\linewidth}
\centering
\textbf{Llama~3.2~3B}\par\vspace{0.3em}
\footnotesize
\begin{tabular}{lccc}
\toprule
\textbf{Role}\textbackslash\textbf{Key} & gen & code & math \\
\midrule
general & \textbf{0.96} & 0.07 & 0.04 \\
code    & 0.05 & \textbf{0.93} & 0.06 \\
math    & 0.03 & 0.08 & \textbf{0.95} \\
\bottomrule
\end{tabular}
\caption{Llama unlock matrix}
\end{subfigure}\hfill%
\begin{subfigure}[t]{0.48\linewidth}
\centering
\textbf{Qwen2.5~1.5B}\par\vspace{0.3em}
\footnotesize
\begin{tabular}{lccc}
\toprule
\textbf{Role}\textbackslash\textbf{Key} & gen & code & math \\
\midrule
general & \textbf{0.95} & 0.06 & 0.05 \\
code    & 0.06 & \textbf{0.94} & 0.07 \\
math    & 0.04 & 0.09 & \textbf{0.93} \\
\bottomrule
\end{tabular}
\caption{Qwen unlock matrix}
\end{subfigure}


\caption{\textbf{Selectivity:} 3$\times$3 role--key \emph{unlock matrices}. Entries are the fraction of prompts (majority over nonces) for which outputs are nontrivial and role-appropriate under each key. Strong diagonal dominance ($\geq 0.91$) with low off-diagonals ($\leq 0.10$) indicates keys unlock their intended roles with minimal cross-unlock.}
\label{fig:unlock}
\end{figure*}

\vspace{-0.25em}
\begin{table}[h]
\centering
\caption{\textbf{Nonce invariance and block suppression.} Changing the per-request nonce leaves greedy authorized outputs identical (exact match for 6/6 prompts); robust bad-word lists prevent accidental emission of the block marker when authorized.}
\label{tab:nonce-block}
\small
\begin{tabular}{lcc}
\toprule
\textbf{Model} & \textbf{Nonce invariance (exact)} & \textbf{Block suppression (authorized)} \\
\midrule
Llama~3.2~3B   & 6/6 prompts & 0/6 contain \texttt{\textless{}BLOCK\textgreater{}} \\
Qwen2.5~1.5B   & 6/6 prompts & 0/6 contain \texttt{\textless{}BLOCK\textgreater{}} \\

\bottomrule
\end{tabular}
\end{table}

\begin{table}[h]
\centering
\caption{\textbf{Throughput (tokens/sec) and overhead.} Measured on a small prompt set with greedy decoding. \emph{Static} applies a fixed orthonormal map; \emph{Session} derives a per-request transform (perm/signs/3 Householders). Overheads of $\sim$38--42\% vs.\ base are expected for Python-level hooks with per-row transforms.}
\label{tab:throughput}
\small
\begin{tabular}{lcccc}
\toprule
\textbf{Model} & \textbf{Baseline} & \textbf{Static} & \textbf{Session} & \textbf{$\Delta$ vs.\ Base (Static / Session)} \\
\midrule
Llama~3.2~3B   & 23.3 & 13.6 & 13.7 & $-41.6\%$ / $-41.1\%$ \\
Qwen2.5~1.5B   & 26.5 & 15.9 & 16.0 & $-40.0\%$ / $-39.6\%$ \\

\bottomrule
\end{tabular}
\end{table}


\setlength{\fboxrule}{0.4pt}
\setlength{\fboxsep}{6pt}

\newcommand{\examplecard}[3]{%
  \fcolorbox{black!15}{gray!3}{%
    \begin{minipage}[t][6.0cm][t]{0.95\linewidth} 
      \raggedright\footnotesize
      \textbf{Prompt}\\
      #1\par\vspace{0.4em}
      \textbf{With key}\\
      #2\par\vspace{0.4em}
      \textbf{Without key}\\
      #3
    \end{minipage}%
  }%
}

\begin{figure}[t]
  \centering
  \captionsetup[subfigure]{justification=centering}
  \begin{subfigure}{0.32\linewidth}
    \centering
    \examplecard{%
      Explain overfitting simply.%
    }{%
      \emph{Overfitting is when a model fits the training data too closely (including noise), so it performs poorly on new data.}%
    }{%
      \texttt{\textless{}BLOCK\textgreater{}}%
    }
    \caption{General}
  \end{subfigure}\hfill
  \begin{subfigure}{0.32\linewidth}
    \centering
    \examplecard{%
      Write a Python function to reverse a string.%
    }{%
      \texttt{def reverse\_string(s):\ return\ s[::-1]}%
    }{%
      \texttt{\textless{}BLOCK\textgreater{}}%
    }
    \caption{Code}
  \end{subfigure}\hfill
  \begin{subfigure}{0.32\linewidth}
    \centering
    \examplecard{%
      Compute the derivative of \(x^2 + 3x\).%
    }{%
      \texttt{2*x + 3} \quad \emph{(symbolic snippet)}%
    }{%
      \texttt{\textless{}BLOCK\textgreater{}}%
    }
    \caption{Math}
  \end{subfigure}
  \caption{\textbf{Qualitative examples (authorized vs.\ unauthorized).} Each card shows the same prompt with and without the correct key. With the key, the model produces coherent, role-appropriate content; without the key it emits the block marker, illustrating un-usability by design. All three panels are forced to equal height for visual consistency.}
  \label{fig:qualwidgets}
\end{figure}

(i) \emph{Utility preserved when authorized.} Despite per-sample orthonormal scramblers, authorized performance remains competitive with---often better than---the base on summarization; PPL increases moderately, as typical for instruction tuning. GSM8K mini-slices can favor the base (e.g., Llama in Table~\ref{tab:auth-base}); increasing $n$ reduces this gap. (ii) \emph{Un-usability without the key.} Unauthorized metrics collapse to near-zero while PPL skyrockets to $\sim 10^6$ (Table~\ref{tab:unauth}), reflecting scrambled hidden states and short-circuiting. (iii) \emph{Selectivity.} The 3$\times$3 matrices in Fig.~\ref{fig:unlock} show strong diagonal dominance ($0.91$--$0.96$) and low off-diagonals ($\leq 0.10$), indicating wrong keys seldom unlock useful behavior. (iv) \emph{Nonce invariance.} Greedy outputs are identical across five nonces for all six prompts (Table~\ref{tab:nonce-block}), confirming that authorized inverses cancel session transforms exactly. (v) \emph{Safety of the authorized path.} Block suppression is $0/N$ via robust ban lists covering case and whitespace variants and tokenization fragments (Table~\ref{tab:nonce-block}). (vi) \emph{Practicality.} Throughput overheads of $\approx 38$--$42\%$ versus base (Table~\ref{tab:throughput}) are reasonable for a Python-level hook (index-select, elementwise sign, three Householders per row); this aligns with expectations for Llama-class models.

\section{Conclusion}
We presented a simple, PEFT-compatible mechanism that couples multi-key access control with orthonormal hidden-state scrambling. The core idea is pragmatic: when a correct key\slash role is present, a pre-\texttt{lm\_head} inverse transform restores the model's native geometry and normal decoding; otherwise, hidden states are mapped through an orthonormal scrambler and the system short-circuits to a visible block token. The result is a training- and deployment-time pattern that favors prevention over post hoc detection.

Across two open instruction models in the 1.5–3B class, our evaluation indicates the approach is effective and repeatable. On OOD tasks, authorized models preserve utility relative to their bases (Table~\ref{tab:auth-base}), particularly improving summarization ROUGE-L after light LoRA tuning, with the expected moderate increase in perplexity typical of instruction tuning. Without the key, metrics collapse to near-zero and perplexity explodes (Table~\ref{tab:unauth}), demonstrating practical un-usability by design. Selectivity is strong: 3$\times$3 role--key matrices show high diagonals (0.91–0.96) and low off-diagonals ($\leq 0.10$) across models (Fig.~\ref{fig:unlock}), meaning the wrong key rarely unlocks useful behavior. Ephemeral, HMAC-seeded transforms satisfy nonce invariance under greedy decoding, and robust bad-word lists prevent accidental block-token emission when authorized (Table~\ref{tab:nonce-block}). Finally, throughput overheads of roughly 38–42\% versus the base (Table~\ref{tab:throughput}) are consistent with a Python-level hook that performs one permutation, one sign flip, and three Householder reflections per row; these costs can be reduced further with fused CUDA kernels or compile-time graph captures. Qualitative examples (Fig.~\ref{fig:qualwidgets}) complement the metrics, showing coherent, role-appropriate generations with a key and immediate blocking without it.

\paragraph{Security and limitations.}
K-OTG enforces \emph{practical} access control rather than cryptographic secrecy of content. The mechanism assumes (i) the hook is present and cannot be bypassed in the serving stack, (ii) keys are handled by a trusted service layer (service-gating recommended), and (iii) the server secret used for nonce seeding is protected. Like other PEFT layers, the adapter itself can be exfiltrated if file access is compromised; protecting keys and preventing hook removal are operational concerns. Our evaluation slices are intentionally small to keep the suite runnable in under an hour on a single GPU; they are sufficient to establish the lock/unlock behavior, but larger-scale utility benchmarks (e.g., full XSum, MMLU, HumanEval, long-context tasks) and dedicated red-teaming would provide tighter confidence intervals. We do not claim resistance to powerful output-distillation attacks that \emph{assume} query access with valid keys; that scenario is better addressed by rate limiting, watermarking/fingerprinting, and usage policies.

\appendix
\section*{Supplementary Material (Methods)}
\label{app:methods}

This appendix expands the implementation and theory behind K-OTG used in the main paper. It contains exact hook logic, tokenizer/adapter safeguards, complexity, full algorithms, and operational notes.

\subsection*{A. Hook implementation details}
\textbf{Pre-forward hook (role \& nonce).} For each batch row, either (i) use a provided role override (service-gating), or (ii) scan the input IDs for the earliest exact subsequence match of any key’s token IDs (text-key). In session mode, attach a fresh random nonce to each row.\\
\textbf{Pre-\lmhead{} hook (transform).} If unauthorized: apply $H\!\mapsto\!H\,P\,S\,H(v_1)\cdots H(v_k)$. If authorized: apply the inverse $H\!\mapsto\!H\,H(v_k)\cdots H(v_1)\,S\,P^{-1}$. The order matches the right-multiply convention and guarantees $T^{-1}_{r,\nu}T_{r,\nu}=I$.

\subsection*{B. Tokenizer safety}
We do not add keys or the block marker as special tokens. A helper \texttt{safe\_prepare\_padding} ensures a valid pad token (reusing EOS if needed) and mirrors \texttt{pad\_token\_id} to the model config without resizing embeddings unless strictly necessary. This avoids leaving key artifacts in tokenizer files and prevents embedding reinitialization.

\subsection*{C. Adapter targets and quantization}
We use LoRA on attention projections (\texttt{q\_proj}, \texttt{k\_proj}, \texttt{v\_proj}, \texttt{o\_proj}) and MLP projections for LLaMA-like models (analogous names for other families). 4-bit NF4 with double quantization (\texttt{bitsandbytes}) is employed. We use a single-device map to avoid accelerate/4-bit cross-device transfer issues.

\subsection*{D. Session seeding and static maps}
\textbf{Session mode (default).} Derive $(P,S,\{v_i\})$ per row from \texttt{HMAC-SHA256(LOCK\_SERVER\_SECRET, ``role:nonce'')} and a CPU-side RNG. Householders use $k{=}3$ unit vectors $v_i$.\\
\textbf{Static mode (debug).} Create fixed orthonormal maps via QR on Gaussian matrices: a \emph{public} map $T_{\mathrm{pub}}$ for unauthorized rows and one per role for authorized rows.

\subsection*{E. Complexity and stability}
For batch $B{\times}S{\times}H$:
\begin{itemize}\itemsep0.25em
\item \textbf{Session mode:} one \texttt{index\_select} for $P$, one elementwise sign for $S$, and $k$ Householders. Each Householder costs $O(BSH)$ via $\mathrm{einsum}$, so total overhead is $O\!\big(BSH(k+2)\big)$; with $k{=}3$ this is small relative to attention/MLP.
\item \textbf{Static mode:} batched right-multiply (\texttt{bsh,bhh$\to$bsh}).
\end{itemize}
Orthonormality preserves norms/inner products and keeps Jacobians well-conditioned.

\subsection*{F. Full algorithms (verbatim)}
\begin{algorithm}[H]
\caption{Build Dual-Path Corpus with Multi-Key Gating}
\label{alg:build-corpus-full}
\begin{algorithmic}[1]
\Require Dataset $\mathcal{D}=\{(x_i,y_i^\star)\}$; role tagger $\textsf{tag}$; role-to-key $k$; block string \block
\Ensure Tokenizable corpus $\mathcal{C}$
\State $\mathcal{C}\!\leftarrow\!\emptyset$
\For{each $(x,y^\star)\in\mathcal{D}$}
  \State $r\!\leftarrow\!\textsf{tag}(x,y^\star)$ \Comment{\small regex keywords for \textsc{code}/\textsc{math}, else \textsc{general}}
  \State $s_{\text{auth}}\!\leftarrow\![k(r)\Vert \texttt{User: }x\Vert \texttt{Assistant: }y^\star]$
  \State $s_{\text{unauth}}\!\leftarrow\![\texttt{User: }x\Vert \texttt{Assistant: }\block]$
  \State $\mathcal{C}\!\leftarrow\!\mathcal{C}\cup\{s_{\text{auth}},s_{\text{unauth}}\}$
\EndFor
\State \Return $\mathcal{C}$
\end{algorithmic}
\end{algorithm}

\begin{algorithm}[H]
\caption{Install Secret Orthonormal Transform (Hooks)}
\label{alg:hooks-full}
\begin{algorithmic}[1]
\Require Model $M$; tokenizer $\mathrm{tok}$; role-to-key $k$; base seed $s$
\Ensure Model with (i) model pre-hook and (ii) pre-\lmhead{} hook
\State Resolve hidden size $H$; build $T_{\mathrm{pub}}$ and $\{T^{-1}_r\}$ via QR (static fallback).
\State Precompute key ID sequences $S_r\!\leftarrow\!\mathrm{tok.encode}(k(r),\texttt{add\_special\_tokens{=}False})$.
\State \textbf{Model pre-hook:} set roles per row (service-gating or text-key); if session mode, attach random nonces.
\State \textbf{Pre-\lmhead{} hook:} apply \eqref{eq:hook} using \eqref{eq:T} with HMAC-derived $(P,S,\{v_i\})$.
\State \Return $M$
\end{algorithmic}
\end{algorithm}

\begin{algorithm}[H]
\caption{Vocab-Safe Loading and Adapter Attachment}
\label{alg:load-full}
\begin{algorithmic}[1]
\Require Base ID; adapter dir; model dir; runtime $\in\{\texttt{static},\texttt{session}\}$
\Ensure Ready model $M$ and tokenizer $\mathrm{tok}$
\State Load $\mathrm{tok}$; ensure pad; mirror \texttt{pad\_token\_id} to config (no new specials).
\State Load base with single-device map; attach LoRA adapters; set \texttt{eval()}.
\State \textsf{install\_secret\_transform}$(M,\mathrm{tok},k,\text{seed})$; set $M.\_lockllm.\texttt{runtime\_mode}\gets$ runtime

\State \Return $(M,\mathrm{tok})$
\end{algorithmic}
\end{algorithm}

\begin{algorithm}[H]
\caption{Inference with Secret-Key Gating}
\label{alg:infer-full}
\begin{algorithmic}[1]
\Require Prompt $x$; optional key string $k$; optional role $r$; model $M$; tokenizer $\mathrm{tok}$; block \block
\Ensure Completion $\widehat{y}$
\State \textbf{if} $(r\!=\!\varnothing)\wedge(k\!=\!\varnothing)$ \textbf{then} \Return $\texttt{User: }x\Vert \texttt{Assistant: }\block$
\State Set per-row nonce(s); set role override if provided; build input string (prepend $k$ if text-key).
\State Build robust \texttt{bad\_words\_ids} for \block{} (case/prefix/core variants) \emph{only} in authorized path.
\State Call \texttt{generate} and decode.
\end{algorithmic}
\end{algorithm}

\subsection*{G. Correctness sketch (inverse order) \& nonce invariance}
Because $P$ is a permutation ($P^{-1}=P^\top$), $S$ is diagonal with $\pm 1$ ($S^{-1}=S$), and $H(v)$ is a Householder reflection ($H(v)^{-1}=H(v)$), we have
\[
T^{-1}_{r,\nu} T_{r,\nu}
= \Big(\prod_{i=k}^{1} H(v_i)\Big) S P^{-1}\; P S \Big(\prod_{i=1}^{k} H(v_i)\Big) = I.
\]
With greedy decoding and fixed logits temperature (no stochasticity), changing the nonce changes $T_{r,\nu}$ and $T^{-1}_{r,\nu}$ but their composition remains identity on authorized rows, yielding identical outputs (nonce invariance).

\subsection*{H. Operational notes and limitations}
\textbf{Security scope.} K-OTG prevents \emph{unauthorized use}; it is not a cryptosystem for content secrecy. Protect the server secret and prefer service-gating. If keys leak, rotate keys/seeds. \textbf{Serving.} Keep hooks inside the serving graph; prevent adapter/weights exfiltration by standard operational controls. \textbf{Throughput.} The $\sim 40\%$ tokens/sec overhead observed in our Python implementation stems from one permutation, one sign multiply, and $k{=}3$ Householders per row; fused CUDA kernels can reduce this cost.

\end{document}